\documentstyle[11pt,fleqn]{article}
\title{Comments on "Structural dynamics and resonance in plants with nonlinear stiffness"} 
\author{A. Kwang-Hua CHU} 
\date{Department of Physics, Xinjiang University,
Urumqi 830046, PR China \\ and \\ P.O. Box 30-15, Shanghai 200030, PR China}
\begin{document}           
\maketitle
\begin{abstract}
We make comments on the paper by  Miller [{\it J. Theor. Biol.}
{\bf 234} (2005) 511]. \newline

\noindent
{\it Keywords:} Resonance; Biomechanics; Biomaterials; Plants; Nonlinear dynamics
\end{abstract}
\doublerulesep=6mm    %
\baselineskip=6mm
\bibliographystyle{plain}               

\noindent
Recently researchers suggested that trees, crops, and other plants often uproot or snap when they are forced
by
gusting winds or waves at their natural frequency (see, e.g., Kerzenmacher and Gardiner, 1998).
This can be attributed to the fact that the deflections of the plant, and hence
mechanical stresses along the stem and root system, are greatest during resonance (cf. Spatz and Speck, 2002).
Motivated by above status, Miller just reported the effects of hardening (elastic modulus
increases with strain) and softening (elastic modulus decreases with strain)
nonlinearities on the structural dynamics of plant stems (Miller, 2005). To better understand the effect
of nonlinear
stiffness on the resonant behavior of plants, Miller modeled plant stems as forced Duffing oscillators
with softening or hardening nonlinearities and found  that the resonant behavior of plants with nonlinear
stiffness is
in the large different from that predicted by linear models of plant structural dynamics.
In the reported results, the maximum amplitudes of deflecton of the plant stem were calculated numerically
for forcing
frequencies ranging from zero to twice the natural frequency. For hardening nonlinearities, the resonant
behavior was 'pushed' to higher frequencies, and the maximum deflection amplitudes were lower than for
the linear case. For softening nonlinearities, the resonant behavior was pushed to lower frequencies,
and the maximum deflection amplitudes were higher than for the linear case. However, Miller found that
damping has the
effect of drastically decreasing deflection amplitudes and reducing the effect of the nonlinearities.
\newline
Some of the details could be introduced below before the present author's comments
Miller modeled the plant as  a system via a Duffing oscillator, a cubic
stiffness term, $k_3 x^3(t)$; is added to the left-hand side of
conventional  linear ordinary differential equation (ODE) describing a
spring-mass-damper system : $m x''(t)+c x'(t)+k_1 x(t)=F(t)$,
where $m$ is the mass, $c$ is the damping coefficient, $k_1$ is
proportional to the stiffness of the spring, and $F(t)$ is the
applied force as a function of time. Here,  $k_3$ is the cubic stiffness coefficient. $k_3 >0$ models a
hardening nonlinearity, and $k_3 <0$ models a softening
nonlinearity.
$x(t)$ gives
the displacement of the stem as a function of time. This
displacement may be described as the horizontal
displacement or angular displacement of a point on
the stem. The choice of $x(t)$ determines the values of the
corresponding constants $c$ and $k_1$. In terms of plant
structural dynamics, the first term of the equation
describes the effective mass of the plant times its
acceleration, the second term describes the friction in
the system due to aerodynamic, material and structural
damping of the plant, the third term describes the
resistance of the trunk or stem to bending, and the
forcing term describes the effective wind force on the
plant. In reality, the damping term is probably not
linearly proportional to velocity. 
Note also that this equation gives motion at one point. \newline
Soft and hard nonlinear structures have often been
modeled as Duffing oscillators (cf. the detailed references in Miller, 2005;
the Duffing oscillator was originally introduced to
model the large amplitude vibration modes of a steel
beam subjected to periodic forces; Duffing, 1918). 
Since most biomaterials have nonlinear stiffness, modeling biological
structures as Duffing oscillators could also be
beneficial in gaining a better understanding of resonance
in plants. Miller explained that the justification for this choice of model is
based on history and simplicity: this is a well-studied
nonlinear model that is relatively simple, lends itself
to some analytical work, and yet has very complex
behavior (Miller, 2005). \newline
The addition of this nonlinear cubic term is
what distinguishes this model from previously proposed
models of plant stem dynamics. An argument can be
made that the Duffing oscillator is the simplest
appropriate nonlinear model available. A linear term
and a nonlinear term are needed to describe the initial
linear-elastic region of the stress-strain curve and the
subsequent nonlinearities under large deformations.
\newline
The following change of variables was made  (Miller, 2005):
$X=x/L_{max}$, $T=\omega_o t$,
where $L_{max}$ is a characteristic maximum deflection, $\omega_o =\sqrt{k_1/m}$ is the natural frequency. In
this case, $L_{max}$ is the deflection at which failure occurs.
\begin{equation}
 X'' (T)+2 \xi X'(T)+X(T)+\frac{K_3}{K_1} X^3 (T)=\frac{p_o}{K_1} F(\frac{\omega}{\omega_o} T),
\end{equation}
here, $\xi=c/(2m \omega_o)$ (the damping ratio), $K_1=k_1 L_{max}$, $K_3=k_3 (L_{max})^3$,
$p_o$ is the forcing amplitude. \newline
Miller's simulations were run for 40 periods of
oscillation. For each set of parameters, 'forward' and
'backward' scans were performed. This technique was
used to ensure that the upper portion of the response
curve was simulated since the solution of the Duffing
equation is dependent upon the initial conditions
(Tufillaro et al., 1992). The forward and backward
scans were performed as follows: during the first 10
oscillations, the forcing frequency was set to a low
frequency ($\omega=0.1 \omega_o$) for the forward scan or a high
frequency ($\omega=2.0 \omega_o$) for the backward scan. During the
next 30 oscillations, the forcing frequency was set to
the forcing frequency considered for the simulation. The
maximum deflection amplitude was taken over the last
20 periods of oscillation. These maximum amplitudes
were plotted as functions of the forcing frequency,
damping ratio, and nonlinear ratio (Miller, 2005).
For a certain range of forcing frequencies, the
response amplitudes become multi-valued. This corresponds
to forcing frequencies that yield more than one
real root (cf.  Eq. (11) in Miller, 2005). In the damped case, the response
amplitudes become multi-valued if the damping is
sufficiently small (cf. Fig. 4 in Miller, 2005). In the multi-valued range of
the graph, the response amplitudes can take on any
value of the branches. The value taken depends upon the
initial conditions. \newline
The present author likes to make some remarks about Miller's interesting paper (Miller, 2005).
Firstly, while Miller used an effective-mass approach, but the plant mass was treated as a fixed quantity which is not realistic
for a plant system since the mass of a living plant is not fixed (there are gain and loss of its mass
 considering the photosynthesis  and the water-evaporation of the leaves and the fluid flowing from the
 soil into the root of the plant). As the wind passes by, once their are stresses or strains to the plant,
 the water content inside the plant will subsequently change and thus the mass of the plant is changing (cf. Chu, 2004)! \newline
Secondly, the plant system is essentially a continuous (dynamical) system which has an infinite degrees of
freedom. As Miller also noticed, asymmetries in the elastic
properties or mass distribution along the stem could
contribute to twisting motions and torsional strains. Thus, coupled twisting and bending could be
significant to the resonant behavior of the plant. Even it was treated as one degree of freedom, the
flow-structure interaction is still complicated
(the plant (structure) disturbs the flow or wind and
there is also a feedback from the flow (say, wind) to the (elastic) structure simultaneously) enough and it
needs to be considered (the inertia, the damping and the stiffness could be altered via this flow-structure
interaction). Meanwhile, the approximation of $x(t)$ (cf. Fig. 1 in Miller, 2005) is not valid once there is larger wind-loading to the plant : the root (network) of the plant will shift its position due to adaptive accomodations which then changes $x(t)$.  In fact, the fixed-end support for the plant could be relaxed to
a support which can slip along the boundary.\newline
Thirdly, although Miller adopted a nonlinear approach for a simple (one degree of freedom) system.
What is the benefit of the Miller's nonlinear approach compared to the linear approach for  several degrees of freedom or continuous (elastic)
system model? As we know, the (nonlinear) anharmonic term in damped Duffing equation (cubic term : $k_3 [x(t)]^3$)
is nothing but to induce a phase shift (Duffing, 1918; Denman, 1998; Vega and Knobloch, 2003) which Miller could numerically obtain and present them in Figures 7 and 9 (cf. Miller, 2005) for hardening and softening nonlinearities. \newline
Fourthly, as Miller noted : {\it Plant structures have a unique capability of regularly
withstanding substantial resonant loading. Similar to the hardening
case, a 'soft' plant might also avoid the effects of
resonance by moving resonance away from the driving
frequency, if the driving frequency is relatively constant}.
But, Miller could only remark that {\it to gain a complete understanding of how plants deal
with resonance and why they fail, further work describing the behavior of plant stems beyond their
linear elastic range is needed. With the use of detailed
data on how the elastic modulus varies with deformation,
more sophisticated models of plant stem dynamics
could be developed. The ODE model of the Duffing
oscillator could also be extended to a continuous partial
differential equation model. Such a model could
describe deflection, mass, and stiffness as a function of
height.} (the last statement is relevant to the second remark the present author made above)
Miller didn't provide the detailed clues how to tune or accommodate the natural noises.
It seems to the present author that the plant subjected to dynamic wind-loading could be a multistable oscillator
(cf. Blekhman and Landa, 2004; Mikhlin and Manucharyan, 2003) as there might be several equilibrium positions.
For example,  using a bistable oscillator described by a Duffing equation as an example, resonances caused by a biharmonical external force with two different frequencies (the so-called vibrational resonances) were considered (Blekhman and Landa, 2004). It was shown that, in the case of a weakly damped oscillator, these resonances are conjugate; they occur as either the low and high frequency is varied. In addition, the resonances occur as the amplitude of the high-frequency excitation is varied. It was also shown that the high-frequency action induces the change in the number of stable steady states; these bifurcations are also conjugate, and are the cause of the seeming resonance in an overdamped oscillator.
 \newline
Finally, it seems the multi-scale (either to the time or to the 1-dimensional space) approach is useful
to the present problem as the spatial-temporal response for the leaves, the stem, and the root (network)
of the entire plant would be rather different (e.g., just to consider the deflection or strain for
the leaves, the stem, and the root of the plant subjected to the wind-loading!).
{\small Acknowledgements. The author is partially supported by the
Starting Funds of 2005-Xinjiang-University-Scholars.}

\end{document}